\documentclass[11pt,a4paper]{article}

\usepackage{amsfonts,amsmath,amssymb}
\usepackage{bm,tensor}
\usepackage{graphicx}
\usepackage{braket}
\usepackage{ifpdf}
\usepackage[utf8]{inputenc}
\usepackage{color}
\usepackage{physics}
\usepackage[colorlinks=true
,urlcolor=blue
,anchorcolor=blue
,citecolor=blue
,filecolor=blue
,linkcolor=blue
,menucolor=blue
,linktocpage=true
,pdfproducer=medialab
,pdfa=true
]{hyperref}

\newcommand{\df}{\text{d}}
\newcommand{\p}{\partial}
\newcommand{\Mpl}{{M_\text{pl}}}
\newcommand{\tx}{\text}
\newcommand{\mc}{\mathcal}

\newcommand{\Msol}{{M_\odot}}

\newcommand{\Mpc}{\mathrm{Mpc}}

\newcommand{\Hz}{\mathrm{Hz}}

\newcommand{\MeV}{\mathrm{MeV}}
\newcommand{\GeV}{\mathrm{GeV}}

\usepackage{cite}

\begin{document}

\begin{titlepage}
\begin{center}

\hfill \today

\vspace{1.5cm}
\baselineskip 8mm

\textbf{\LARGE
Gravitational waves from type~II axion-like curvaton model and 
its implication for NANOGrav result
}

\vspace{2.0cm}
\textbf{\large Masahiro Kawasaki}$^{(a,b)}$
and
\textbf{\large Hiromasa Nakatsuka}$^{(a)}$

\vspace{1.0cm}
\textit{%
$^{(a)}${ICRR, The University of Tokyo, Kashiwa, Chiba 277-8582, Japan}\\
$^{(b)}${Kavli IPMU (WPI), UTIAS, The University of Tokyo, Kashiwa, Chiba 277-8583, Japan}
}

\vspace{2.0cm}
\abstract{


The recent report of NANOGrav is gathering attention since its signal can be explained by a stochastic background of gravitational waves (GWs) with an abundance of $\Omega_{\rm GW}\sim 10^{-9}$ at the reference frequency $f\sim 10^{-8}\Hz$.
The PBH formation scenario is one of the candidates for the NANOGrav signal, which can simultaneously explain the observed $30\Msol$ black holes in the binary merger events in LIGO-Virgo collaboration.
We focus on the type-II axion-like curvaton model of the PBH formation.
In the type~II model, the complex field whose phase part is the axion rolls down from the origin of the potential.
It is found that type II model achieves the broad power spectrum of the density perturbations and can simultaneously explain the LIGO-Virgo events and the NANOGrav signal.
We also improve the treatment of the non-Gaussianity of perturbations in our model to accurately estimate the amplitude of the induced GWs.

}
\end{center}
\end{titlepage}
\setcounter{footnote}{0}

\baselineskip 6mm

\section{Introduction}
\label{sec_Introduction}

North American Nanohertz Observatory for Gravitational Waves (NANOGrav) is one of the Pulsar-timing-array experiments, which has been trying to detect the gravitational wave (GW) signal through the long-term observation of pulsars.
Recently the NANOGrav reported a hint of the stochastic GW signal in their 12.5--year observation~\cite{Arzoumanian:2020vkk}.
This signal can be explained by the stochastic GW with $\Omega_{\rm GW}\sim 10^{-9}$ at $f\sim 10^{-8}\Hz$, which can be sourced by cosmic string~\cite{Ellis:2020ena,Blasi:2020mfx,Buchmuller:2020lbh,Samanta:2020cdk,Datta:2020bht,Ramberg:2020oct}, phase transitions~\cite{Nakai:2020oit,Addazi:2020zcj,Neronov:2020qrl,Li:2021qer,Barman:2020jrf}, large density fluctuations in primordial black hole (PBH) formation models~\cite{Vaskonen:2020lbd,DeLuca:2020agl,Inomata:2020xad,Kohri:2020qqd,Domenech:2020ers,Sugiyama:2020roc,Braglia:2020eai,Bhaumik:2020dor,Braglia:2020taf,Atal:2020yic} and other origin \cite{Vagnozzi:2020gtf,Bhattacharya:2020lhc,Kuroyanagi:2020sfw}.

The PBH formation scenario is an attractive candidate for the NANOGrav signal since it can simultaneously explain the observed $30\Msol$ black holes in the binary merger events in LIGO-Virgo collaboration \cite{LIGOScientific:2018jsj,Bird:2016dcv,Clesse:2016vqa,Sasaki:2016jop}.
The PBH formation requires the large curvature power spectrum ($\mathcal P_\zeta(k_{\rm pbh})\sim 0.02$) at a wavenumber $k_{\rm pbh}$ related to the mass of the formed PBHs.
Since PBHs are formed when the high-density regions enter the horizon and collapse, the PBH mass is roughly given by the horizon mass at the collapse, which is related to the frequency of the density fluctuation $f$ ($=k_\text{pbh}/2\pi$) as~\cite{Inomata:2017vxo}
\begin{align}
     M(f)
&\simeq
	30 \Msol 
	\left( \frac{\gamma}{0.2}  \right)
	\left( \frac{\tx{g}_*}{10.75}  \right)^{-1/6}
	\left( \frac{f}{5.3\times 10^{-10}\,\Hz}  \right)^{-2},
	\label{PBHmasses}
\end{align}
where $\gamma\sim 0.2$ is a ratio of PBH mass to the horizon mass~\cite{1975ApJ...201....1C}
and $\tx{g}_*$ is the numbers of relativistic degrees of freedom at the PBH formation.
The same density fluctuations that collapse to form PBHs also generate a stochastic background of GWs through the nonlinear coupling, $\Braket{\zeta\zeta h}$ when they enter the horizon~\cite{Saito:2008jc,saito2010gravitational,Unal:2020mts}, with a spectrum peaking at the frequency $f\sim 10^{-9}\Hz$, very close to the one of the NANOGrav signal.

In this paper, we focus on the axion-like curvaton model of the PBH formation scenario \cite{Kawasaki:2012wr,Kawasaki:2013xsa,Ando:2017veq,Inomata:2020xad,Ando:2018nge}.
There are two types of the axion-like curvaton models;  in one type the complex field $\Phi$ (whose phase part is the axion) rolls down toward origin during inflation ~\cite{Kawasaki:2012wr,Kawasaki:2013xsa,Ando:2017veq,Inomata:2020xad}, and in the other type $\Phi$ rolls down from the origin~\cite{Ando:2018nge}.
These two types lead to the different dynamics and the power spectrum of induced GWs. 
The former type (type I) has been already studied and successfully explains the NANOGrav signal~\cite{Inomata:2020xad}.
The type~II model was first proposed in~\cite{Ando:2018nge} where the model parameters were chosen to achieve a narrow density power spectrum of the density perturbations.
However, a broader power spectrum is preferable to account for the NANOGrav signal.
The goal of this paper is to investigate if such a broad power spectrum can be produced also in the type II model.
It is found that by choosing appropriate parameters of the potential term in Eq.~\eqref{eq_params}, we obtain the desired broader spectral shape. 
Moreover, since our model produces large positive non-Gaussianity on the density power spectrum, we can achieve the required abundance of PBHs by the smaller amplitude of the density power spectrum than the Gaussian one.
We improve the treatment of the non-Gaussianity of this model to accurately estimate the required amplitude of density power spectrum, and also the induced GW.

This paper is organized as follows.
We calculate the curvature power spectrum of the type II axion-like curvaton model in Sec.~\ref{sec_curvatonmodel}.
Next, we calculate the PBH abundance in Sec.\ref{sec_pbhformation} and induced GWs in Sec.~\ref{sec_gw}.
We can achieve the broad power spectrum of induced GWs and explain the NANOGrav signal.
In Sec.~\ref{sec_conclusion}, we conclude this paper.

\section{Type~II axion-like curvaton model}
\label{sec_curvatonmodel}

{We now briefly summarize the dynamics of background and fluctuations in}
the type II axion-like curvaton model. (See \cite{Ando:2018nge} for the detailed calculation.)
In this model, large fluctuations are produced in the phase direction of a complex scalar field $\Phi$, which we call ``axion-like curvaton''.
The potential of $\Phi$ is given by~\cite{Ando:2018nge}
\begin{align}
	V_{\Phi}=\frac{\lambda}{4} 
	\left(
		|\Phi|^2
		-\frac{v^2}{2}
	\right)^2 
	+g I^2 |\Phi|^2
	- v^3\epsilon(\Phi+\Phi^*),
	\label{potecomp}
\end{align}
where $I$ is the inflaton field, $\lambda$ and $g$ are coupling constants, $v/\sqrt{2}$ is the vacuum expectation value after inflation.
The bias term, $v^3\epsilon(\Phi+\Phi^*)$, is introduced to avoid the cosmic string problem and the stochastic effect on dynamics of $\Phi$, which requires the complicated treatment of fluctuations.
Note that $\epsilon$
is naturally small ($\epsilon \ll 1$) in the sense of 't Hooft's naturalness~\cite{tHooft:1979rat} since $U(1)$ symmetry is restored for $\epsilon =0$.

The field value of $\Phi$ changes during inflation due to the first and second term of Eq.~\eqref{potecomp}.
In the early stage of inflation ($I \gtrsim (\lambda/g)^{1/2}v$), the interaction term with the inflaton fixes $\Phi$ near the origin.
In the late stage, the inflaton field value becomes smaller and $\Phi$ starts to roll down the Higgs-like potential towards $|\Phi|=v$.
In this dynamics the phase direction acquires large fluctuations $\sim H/|\Phi|$ ($H$: Hubble parameter during inflation).

The PBH formation requires the large fluctuations with wavenumber $k=k_{\rm pbh}\simeq 10^5 \,\Mpc^{-1}$, which corresponds to the scale of $30\Msol$ PBHs. 
Thus, we determine the inflaton coupling $g$ so that  $\Phi$ starts to roll down potential at $t_{\rm pbh}$ satisfying  $H(t_\text{pbh})=k_\text{pbh}/a(t_\text{pbh})$, which leads to
\begin{align}
	g\simeq
	\frac{\lambda}{4}
	\left(
	\frac{v}{I(t_{\rm pbh}) }
	\right)^2.
	\label{eq_gasum}
\end{align}
We also assume that the effective mass of $\Phi$ is much larger than the Hubble parameter ($\lambda v^2\gg H^2$) until $t_{\rm pbh}$ to suppress the fluctuation at the larger wavelengths.
This condition also ensures independence on the initial condition since $\Phi$ settles near the origin of the potential independently of the initial field value of $\Phi$.

Based on the above background field dynamics, we calculate the fluctuation of $\Phi$, which is decomposed as 
\begin{align}
	\Phi=\frac{1}{\sqrt{2}}(\varphi_0+\varphi) e^{i\theta }
\end{align}
with the homogeneous solution $\varphi_0$, and the perturbations $\varphi$ and $\theta$.
The phase direction $\theta$ works as the curvaton, and the canonical field of the phase direction is defined as $\tilde \sigma \equiv  \varphi_0 \theta$.
The inflation induces the quantum fluctuations with amplitude $H/(2\pi)$ for $\tilde \sigma$ at the horizon crossing. 
Thus, the power spectrum of the $\theta$ fluctuations  is given by
\begin{align}
	\mathcal P_{\theta}(k,t_k)
	= \left(
	\frac{H(t_k)}{2\pi \varphi_0(t_k)}
	\right)^2 ,
	\label{anguHubble}
\end{align}
where $t_k$ is the time when the fluctuation with $k$ crosses the horizon.
$\mathcal P_{\theta}(k,t_k)$ is suppressed for $k>k_\tx{pbh}$ since the $\varphi_0(t_k)$ quickly grows after $t_\tx{pbh}$.
While the $\Phi$ is fixed near the origin before $t_\text{pbh}$,  $\tilde \sigma$ has the large effective mass given by
\begin{align}
		\tilde{m}_\sigma^2&\equiv 
		\frac{\p^2 \left(	-v^3 \epsilon (\Phi+\Phi^*)\right)}{ \p\tilde \sigma^2}
		\bigg| _{\tilde \sigma=0}
		=
		{\sqrt{2} \epsilon v^2\frac{v}{\varphi_0} } ,
		\label{msigma}
\end{align}
where we choose the model parameter so as $m_{\tilde \sigma }^2 >H^2$ for $t<t_\tx{pbh}$. 
Such large effective mass damps the fluctuations as $\tilde \sigma \propto a^{-3/2}$, and we define the damping factor of the phase direction with momentum $k$  as
\begin{align}
	R_k&
	\equiv
	\left(
	\frac{\tilde \sigma_k(t_\tx{pbh}) }{\tilde \sigma_k(t_k)}
	\right)
	\sim
	\left( \frac{k}{k_\text{pbh}}  \right)^{3/2}
	\text{  for   }
	k<k_\tx{pbh}.
	\label{Dampsim}
\end{align}
($R_k =1$ for $k> k_\text{pbh}$.)
Finally, the power spectrum of the $\theta$ fluctuations at the end of inflation $t_e$ is given by 
\begin{equation}
    P_\theta(k,t_e) = R_k^2 \mathcal P_\theta(k,t_k).
\end{equation}

Let us evaluate the density perturbations induced by the fluctuations of the phase direction.
We suppose that the curvaton obtains the axion-like potential after inflation through some nonperturbative effect.
The potential minimum of the nonperturbative term does not coincide with that of the primordial one determined by the bias term in our model.
Suppose that nonperturbative potential takes the minimum at $\theta=\theta_i$, then it is written as %
\begin{align}
	V_\sigma
	=\Lambda^4
	\left[
	    1-\cos\left( \Theta	 \right)
	\right]
	\simeq 
	\frac{1}{2}m_\sigma^2 \sigma^2 ,
	\label{eq:axion_pot}
\end{align}
where $\Theta \equiv\theta-\theta_i$, the curvaton $\sigma$ is defined as $\sigma \equiv v\Theta$ and $m_\sigma=\Lambda^2/v$.
We assume that $m_\sigma$ is small enough to neglect the axion-like potential during inflation, i.e. $ H^2 \gg m_\sigma^2$.
The density fluctuation is given by 
${\delta \rho_{\sigma}}/{\rho_\sigma}=2\delta\theta/\theta_i$.

Neglecting a small contribution from the inflaton  ($\mathcal{P}_\zeta \sim 10^{-10}$) compared to the curvaton, the power spectrum of curvature perturbations is given by
\begin{align}
\mc P_{\zeta}(k)
    &=
	\left( \frac{r}{4+3r} \right)^2
	\left( \frac{2}{\theta_i} \right)^2
	\mc P_{\theta}(k,t_k)
	R_k^2
	,
	\label{Pzetaformula}
\end{align}
where $r$ is the ratio of the energy density of the curvaton to that of the inflaton (or radiation after reheating), $r=\rho_\sigma/\rho_I$.
Assuming the instant reheating at $t_{\rm reheat}$, the ratio is given by $r( t_{\rm reheat} ) =(v^2\theta_i^2)/(6M_{pl}^2)$, which is chosen to be small to ensure that $\Phi$ does not disturb the inflation.
$r$ grows during radiation-dominated era as $r\propto a$ due to the matter-like behavior of the curvaton, and its growth ends at the curvaton decay, $t_{\rm decay}$.
We require that the curvaton decays into radiation before it overcloses the universe, $r(t_{\rm decay})\sim  0.5 $, at which the temperature of the universe is $T_{\rm decay}\sim (r( t_{\rm reheat} ) /r( t_{\rm decay} ) ) T_{\rm reheat}$.
The typical decay rate of $\Phi$ is related to the curvaton mass as $\Gamma_\sigma = \kappa^2 m_\sigma^3/(16\pi v^2)$ where $\kappa$ is a coupling constant.
We confirm that $T_{\rm decay}\sim 10^3\GeV$ is achieved for $m_\sigma\sim 10^8\GeV$ for our parameters in Eq.~\eqref{eq_params}.
In the following, $r$ refers to the energy ratio after decay, that is $r\equiv r(t_{\rm decay})$.

The PBH abundance highly depends on the eventual non-Gaussian distribution of $\mc P_{\zeta}(k)$.
It is known that the axion-like curvaton models produce large non-Gaussianity with local type bispectrum characterized by the following parameter: \cite{Ando:2017veq}
\begin{align}
    f_{\rm NL} = \frac{5}{12}\left( 
    -3+\frac{4}{r}
    +\frac{8}{4+3r}
    \right).
    \label{eq_fnl}
\end{align}
We discuss the enhancement of the PBH abundance by non-Gaussianity in Sec.\ref{sec_pbhformation}.
Non-Gaussianity also affects the power spectrum of the curvature perturbations and induced gravitational wave through the higher-order correlations \cite{Unal:2018yaa,Cai:2018dig}, whose  contribution is characterized by $\mathcal P_\zeta f_{\rm NL}^2$.
We take non-Gaussianity into account only approximately 
since the Gaussian contribution is the dominant for our choice of parameters
given by Eq.~\eqref{eq_params}, $\mathcal P_\zeta f_{\rm NL}^2<0.04$.
We estimate non-Gaussian amplification on the curvature power spectrum as
\begin{align}
    &Q^{\rm (NL)}_{\mathcal P}
    \equiv
    \frac{\mathcal P_\zeta^{\rm (NL)}(k_*)}{\mathcal P_\zeta(k_*)}
    \nonumber
\\&    =
    1+ 
    \left(
    \frac{3}{5}f_{\rm NL}
    \right)^2
    \frac{k_*^3}{2\pi \mathcal P_\zeta({\bm k_*})}
    \int \df ^3 q \frac{
        \mathcal P_\zeta({\bm q})\mathcal P(|{\bm k_*-\bm q}|)
    }{q^3|\bm k_*-\bm q|^3},
    \label{eq_Q_P}
\end{align}
where $k_*$ is the wavenumber at the peak of $\mathcal P_\zeta$.
$Q^{\rm (NL)}_{\mathcal P}$ is about $1.07$ for $r=0.5$ and $1.01$ for $r=1.0$ in our parameter set.

Finally, the power spectrum of the density perturbations is given by
\begin{align}
    	\mc P_{\delta}(k,t)
	&=
	\left( 
	\frac{2}{3}\frac{k}{a(t)H(t)}
	\right)^4
	T(k\eta(t))^2
	\mc P_{\zeta}(k),
\end{align}
where $\eta(t)$ is the conformal time and $T(x)$ is the transfer function during radiation dominated era, which includes the suppression of the density perturbation in sub-horizon as
\begin{align}
    T(x)\equiv&~
    3\frac{\sin(x/ \sqrt{3})-(x/ \sqrt{3})\cos(x/\sqrt{3})}{ (x/\sqrt{3})^3 } .
    \label{eq_transf}
\end{align}

\begin{figure}[t]
    \centering
    \includegraphics[width=.80\textwidth ]{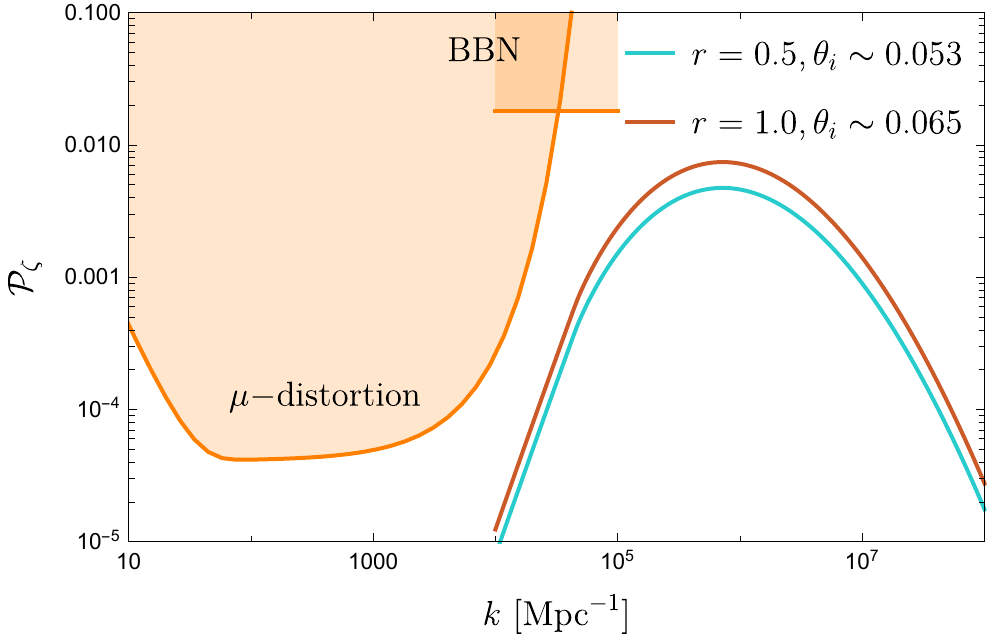}
    \caption{
    The curvature power spectrum $Q^{\rm (NL)}_{\mathcal P} \mathcal P_\zeta(k) $ based on Eqs.\eqref{Pzetaformula} and \eqref{eq_Q_P}.
    The non-Gaussian contribution is included by $Q^{\rm (NL)}_{\mathcal P}$.
    The orange regions are constraints on the curvature power spectrum from $\mu$--distortion \cite{0004-637X-758-2-76} by COBE/FIRAS \cite{Fixsen:1996nj} and BBN~\cite{Inomata:2016uip}. 
    }
    \label{fig_pzeta}
\end{figure}

In this paper, we use the following set of model parameters:
\begin{align}
	\lambda  &= 7.5\times 10^{-6}
	,\quad
	v = 5.77\times 10^{-2}\Mpl
	,\quad \nonumber
\\	g&= 9.56\times 10^{-11}
	,\quad
	\epsilon = 2.57\times 10^{-10} ,
	\nonumber
\\
    &\begin{cases}
        &r= 0.5,\quad \theta_i = 5.3\times 10^{-2}
        \\
        &r= 1.0,\quad \theta_i = 6.5\times 10^{-2}
    \end{cases}
    ,
	\label{eq_params}
\end{align}
where we take $r$ and $\theta_i$ to achieve the PBH abundance required to explains the LIGO-Virgo event rate.
Here we remark some relations between the parameters and the shape of the power spectrum.
The larger $r$ or smaller $\theta_i$ increases the abundance of ALP as Eq.~\eqref{Pzetaformula}, and smaller $\epsilon$ changes the dynamics of $\varphi_0$ at $t\sim t_{\rm pbh}$, both of which result in larger amplitude of the power spectrum.
The peak wavenumber of the power spectrum is determined by the ratio of potential terms, $g/(\lambda v^2)$, as discussed in Eq.~\eqref{eq_gasum}.
The width of the spectrum depends on how slowly $\varphi_0$ changes since the fluctuation of $\theta$ with mode $k$ depends on the field value $\varphi_0$ at the horizon crossing as Eq.~\eqref{anguHubble}.
We flatten the potential by choosing the smaller $v$, $g$ and $\lambda$ compared to the previous paper~\cite{Ando:2018nge} to achieve a broad power spectrum.
We numerically evaluate the dynamics of $\Phi$ assuming the chaotic inflation whose potential is given by $V_I=m_I^2 I^2/2$ with $m_\phi \simeq 10^{13}\GeV$.
The similar dynamics also holds for other inflation models. 
We show the curvature power spectrum $\mathcal P_\zeta(k)$ [Eq.~\eqref{Pzetaformula}] in Fig.~\ref{fig_pzeta}.
We also show the constraints on $\mu$--distortion \cite{0004-637X-758-2-76} by COBE/FIRAS \cite{Fixsen:1996nj} and BBN~\cite{Inomata:2016uip}, and our model safely avoids the current constraints.

\section{PBH formation}
\label{sec_pbhformation}

The PBHs are formed by the collapse of high-density regions when they re-enter the horizon.
The criterion on the PBH formation is estimated by the numerical simulation \cite{Shibata:1999zs}, in which the threshold value of the averaged density fluctuation is obtained.
Since the threshold value is too large to neglect the nonlinear contribution, we need to use the effective threshold value including the nonlinear relation between density and curvature perturbations.
The detailed analysis show that the effective threshold value is $\delta_{\rm th(eff)}\simeq \sqrt{2}\times 0.53$ for the averaged linear density perturbation defined by \cite{Kawasaki:2019mbl}
\begin{align}
	\bar \delta_R (\bm x)
&\equiv 
	\int\df^3 y 
	W(|\bm x-\bm y|,R)
	\delta(\bm y)
	\nonumber
\\&	=
	\int \frac{\df^3 p}{(2\pi)^3}
	\tilde W(p R)e^{i\bm p\cdot \bm x}\delta_p,
	\label{eq:deltaW}
\end{align}
where $R=k^{-1}$ is the scale of the horizon corresponding to the PBH mass [Eq.\eqref{PBHmasses}], $W(x,R)$ and $\tilde W(z)$ are the window functions in the real and Fourier spaces,
and the $\delta_p$ is the density fluctuation.
Although it is known that the choice of window function causes a large uncertainly~\cite{Ando:2018qdb}, 
a natural choice, often used in the literature, is  
the real-space top-hat window function used in the threshold value in the numerical simulation,
\begin{align}
\tilde W(z)
&=
    3\frac{\sin(z)-z\cos(z)}{z^3}.
\end{align}
Thus, a PBH is formed when a region has the averaged density $\bar{\delta}_R$ larger than the threshold value $\delta_{\rm th(eff)}$.

We estimate the PBH formation rate based on the Press--Schechter formalism, where the PBH formation rate is calculated by the probability distribution of the averaged density perturbations.
In our model, $\bar \delta_R$ follows the non-Gaussian distribution due to $f_{\rm NL}$ given by Eq.~\eqref{eq_fnl}, which drastically changes the PBH abundance.
The probability distribution of $\bar \delta_R$ is characterized by the variance and skewness defined by 
\begin{align}
		\sigma_R^2&= 
		\Braket{\bar\delta_R^2}-\Braket{\bar \delta_R}^2,
		\label{eq:variancedef}
		\\
		\mu_R
		&=\sigma_R^{-3}
		\left(
		\Braket{\bar \delta _{R}^3}
		-3\Braket{\bar \delta _R^2}
		\Braket{\bar\delta _R}
		+2\Braket{\bar \delta _R}^3
		\right),
		\label{eq:skewdef}
\end{align}
where $\Braket{...}$ describes the ensemble average of $\delta_p$.
For simplicity, we neglect the scale dependence of $\mu_R$ and evaluate it at $R= R_{\rm pbh}=k_{\rm pbh}^{-1}$, which corresponds to $30\Msol$ PBHs.
Using the formula in \cite{Kawasaki:2019mbl}, the skewness is numerically given by
\begin{align}
    \mu\equiv \mu_R|_{R=R_{\rm pbh}}
    \simeq 3.39f_{\rm NL}\sigma_R|_{R=R_{\rm pbh}}
    .
\end{align}

We construct the statistical variable which reproduces the probability distribution of  $\bar \delta_R$.
Using the Gaussian variable $\chi$ characterized by $\Braket{\chi^2} = \sigma_R^2$, we define the statistical variable as
\begin{align}
		\bar \delta[\chi]
		\equiv \chi
		+\frac{
			\mu
		}{6\sigma_R}
		(\chi^2-\sigma_R^2)  ,
		\label{eq:delhcapprox}
\end{align}
which has the same variance and skewness in Eqs.~\eqref{eq:variancedef} and \eqref{eq:skewdef} up to $\mathcal O(f_{\rm NL}\sigma_R)$.
The probability distribution of $\bar \delta[\chi]$ is given by
~\cite{Young:2013oia,Byrnes:2012yx,Young:2015cyn}
\begin{align}
		P_{R}^{\rm (NG)}(\bar \delta)
		&=\sum_{i=\pm}
		\left| \frac{\df \chi_{i}(\bar \delta)}{\df \bar \delta}\right|
		P_{R}^{\rm (G)}(\chi_{i}(\bar \delta))
		,
\end{align}
where $P_{R}^{\rm (G)}(\chi)$ is the Gaussian distribution function,
\begin{align}
	P_{R}^{\rm (G)}(\chi)
	=\frac{1}{\sqrt{2\pi}\sigma_R } 
	\exp\left(
	-\frac{1}{2} \frac{\chi^2}{\sigma_R^2}
	\right),
\end{align}
and $\chi_\pm(\bar \delta)$ are two solutions of $\bar \delta = \bar \delta[\chi]$,
\begin{align}
		\chi_\pm (\bar \delta)
		=\frac{3\sigma_R}{\mu}
		\left(
		-1\pm \sqrt{
			1+\frac{2\mu}{3} 
			\left(  \frac{\mu}{6}  +  \frac{\bar \delta}{\sigma_R } \right)
		}
		\right).
\end{align}

The PBH formation rate is given by the probability of $\bar \delta > \delta_{\rm th(eff)}$, which leads to
\begin{align}
	\beta(R)
&=
	\int_{\bar \delta>\delta_{\rm th(eff)}} 
	P_{R}^{\rm (NG)}(\bar \delta) \text{d} \bar \delta
=
    \int_{\bar \delta[\chi]>\delta_{\rm th(eff)}} 
    	P_{R}^{\rm (G)}(\chi) \text{d} \chi
    \nonumber
\\&\simeq 
		\frac{\sigma_R }{\sqrt{2\pi} \chi_+(\delta_{\rm th(eff)} ) }
		\exp\left( -\frac{   \chi_+(\delta_{\rm th(eff)} ) ^2  }{2\sigma_R^2} \right).
	\label{eq:probnongau3}
\end{align}
Here we have used $\mu>0$ and $ \chi_+(\delta_{\rm th(eff)} ) /\sigma_R  \gg 1$ in the last line.
The present PBH abundance is given by
\begin{align}
    \label{eq:massspec}
&    f(M)
\equiv
    \frac{\df \Omega_\tx{PBH} }{\df \ln M} 
    \frac{1}{\Omega_\tx{DM}}
\\&	=
    \frac{\beta(R(M))}{1.8 \times 10^{-8}}
    \left( \frac{\gamma}{0.2} \right)^{3/2}
    \left( \frac{10.75}{\tx{g}_{*}} \right)^{1/4}
    \left( \frac{0.12}{\Omega_\tx{DM}h^2} \right)
    \left( \frac{M}{ \Msol } \right)^{-1/2}
	,
\end{align} 
where $\tx{g}_*$ is the number of relativistic degrees of freedom at $T\sim 30\MeV$.

\begin{figure}[t]
    \centering
    \includegraphics[width=.80\textwidth ]{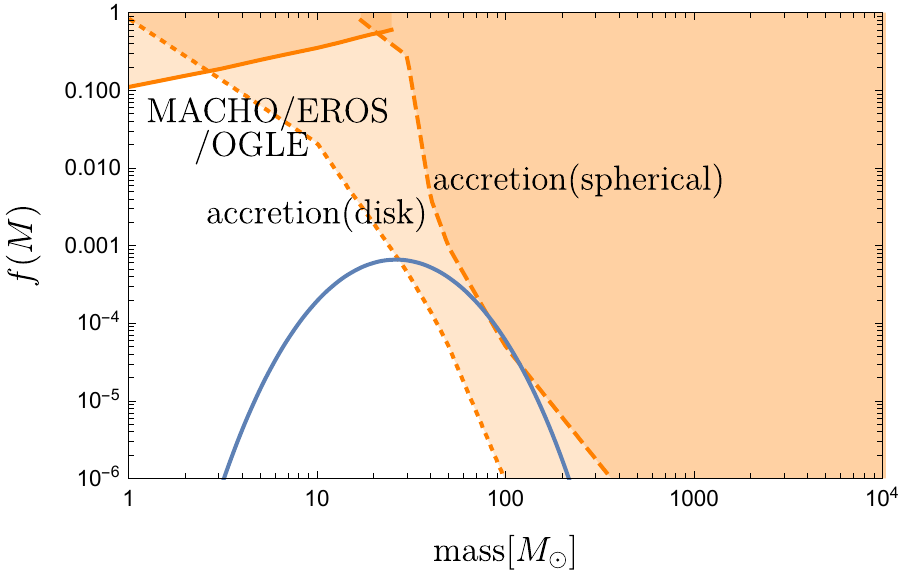}
    \caption{
    The mass spectrum of PBHs (blue line) and the constraints (orange regions) by  microlensing experiments ``MACHO/EROS/OGLE''(solid line)~\cite{Allsman:2000kg,Tisserand:2006zx,Wyrzykowski:2011tr},
    CMB through spherical accretion (dashed line) and disk accretion (dotted line) \cite{Serpico:2020ehh}.
    }
    \label{fig_pbhdist}
\end{figure}

We show the calculated mass spectrum of PBH in Fig.~\ref{fig_pbhdist}.
We also plot the relevant constrains by microlensing experiments ``MACHO/EROS/OGLE''~\cite{Allsman:2000kg,Tisserand:2006zx,Wyrzykowski:2011tr} and energy injection into CMB through accretion around PBHs~\cite{Serpico:2020ehh}.
Since our model predicts the broad mass distribution, the large mass part of the distribution could conflict with the accretion constraints.
However, it is noticed that accretion constraint has a large uncertainty.
In fact, two different constraints are obtained depending on assumptions as shown in Fig.~\ref{fig_pbhdist}.
Our mass distribution is consistent if we adopt the weaker accretion constraint.

\section{Induced gravitational waves}
\label{sec_gw}

The large density fluctuations induce the gravitational waves through the nonlinear coupling $\Braket{\zeta\zeta h}$ when they re-enter the horizon.
The current energy fraction of GWs is written as
\begin{align}
\Omega_{\rm GW} (t_{0},k)
	&=	
	\left(         
		\frac{a_c^2 H_c}{a_0^2 H_0}
	\right)^2
	\Omega_{\rm GW}(\eta_c,k)
	\nonumber
\\&\simeq 
	0.83
	\left( \frac{\tx{g}_c}{10.75} \right)^{-1/3}
		\Omega_{r,0}
		\Omega_{\rm GW} (t_c,k),
		\label{eq_GW_def}
\end{align}
where $\Omega_{r,0}$ is the current energy fraction of radiation, the subscript ``c'' denotes values when GW production effectively finishes.

The energy density of the induced GWs at $t_c$ is calculated by solving the equation of motion of GWs with the source term of scalar fluctuations, and it is given by~\cite{Inomata:2017vxo}
\begin{align}
&\Omega_{\rm GW}(t_c,k)
=
\frac{8}{243}
	\int^\infty_0\df y
	\int^{1+y}_{ \left|1-y\right| }\df x
	 \mc P_\zeta(kx)\mc P_\zeta(ky)
	\frac{y^2}{x^2}
\nonumber
\\&	\qquad\times
    \left(  1- \frac{(1+y^2-x^2)^2}{4y^2}  \right)^2
	\overline{\mathcal I(x,y,\eta_c)}^2
      ,
\end{align}
where overline means the time average over $\eta_c$.
$\mathcal I(x,y,\eta_c)$ is written as
\begin{align}
    \mathcal I(x,y,\eta_c) = 
		\frac{k^2}{a(\eta_c)}
			\int^{\eta_c}\df \bar\eta
			a(\bar\eta)
			g_k(\eta_c;\bar\eta)
			f(ky,kx,\bar\eta)
			.
\end{align}
Here $g_k$ is the Green function,
\begin{align}
    g_k(\eta,\tilde\eta)
	&=
	\frac{\sin(k(\eta-\bar\eta)) } {k}
	\theta (\eta-\bar\eta)    ,
\end{align}
and $f(k_1,k_2,\eta)$ is given by
\begin{align}
    f(k_1,k_2,\eta)
    &=
    \bigg[
    2T  (k_1,\eta)T(k_2,\eta)
    \nonumber
    \\
     +&\left(  
    	\frac{\dot T(k_1,\eta)}{H(\eta)}
    	+T(k_1,\eta)
    \right)
    \left(
    	\frac{\dot T(k_2,\eta)}{H(\eta)}
    	+T(k_2,\eta)
    \right)
    \bigg],
\end{align}
where $T(x)$ is the transfer function of the scalar fluctuations given by Eq.\eqref{eq_transf}.

We comment on the non-Gaussian contribution on the induced gravitational waves~\cite{Garcia-Bellido:2017aan,Unal:2018yaa,Cai:2018dig}.
It is pointed out in~\cite{Cai:2018dig} that non-Gaussianity of scalar fluctuations amplifies the induced  gravitational waves when $\mathcal P_\zeta f_{\rm NL}^2$ is large.
Since $\mathcal P_\zeta f_{\rm NL}^2<0.04$ in our calculation, the effect of non-Gaussianity is expected to be sub-dominant.
Thus, we approximately include the effect of the non-Gaussianity on $\Omega_{\rm GW}$  by multiplying the factor $(Q^{\rm (NL)}_{\mathcal P})^2$ given by  Eq.~\eqref{eq_Q_P}.
This approximation includes a part of non-Gaussian contributions, 
``Hybrid'' and ``Reducible''-type terms discussed in~\cite{Unal:2018yaa}.
Hybrid type is a product of the Gaussian and non-Gaussian contribution of curvature perturbation, and Reducible type is that of non-Gaussian and non-Gaussian contribution.
Although there are other types of sources of GWs, it is known that Hybrid-type is one of the largest contributions among them. 
Thus, our calculation can estimate most of the effects of non-Gaussianity.

\begin{figure}[t]
    \centering
    \includegraphics[width=.90\textwidth ]{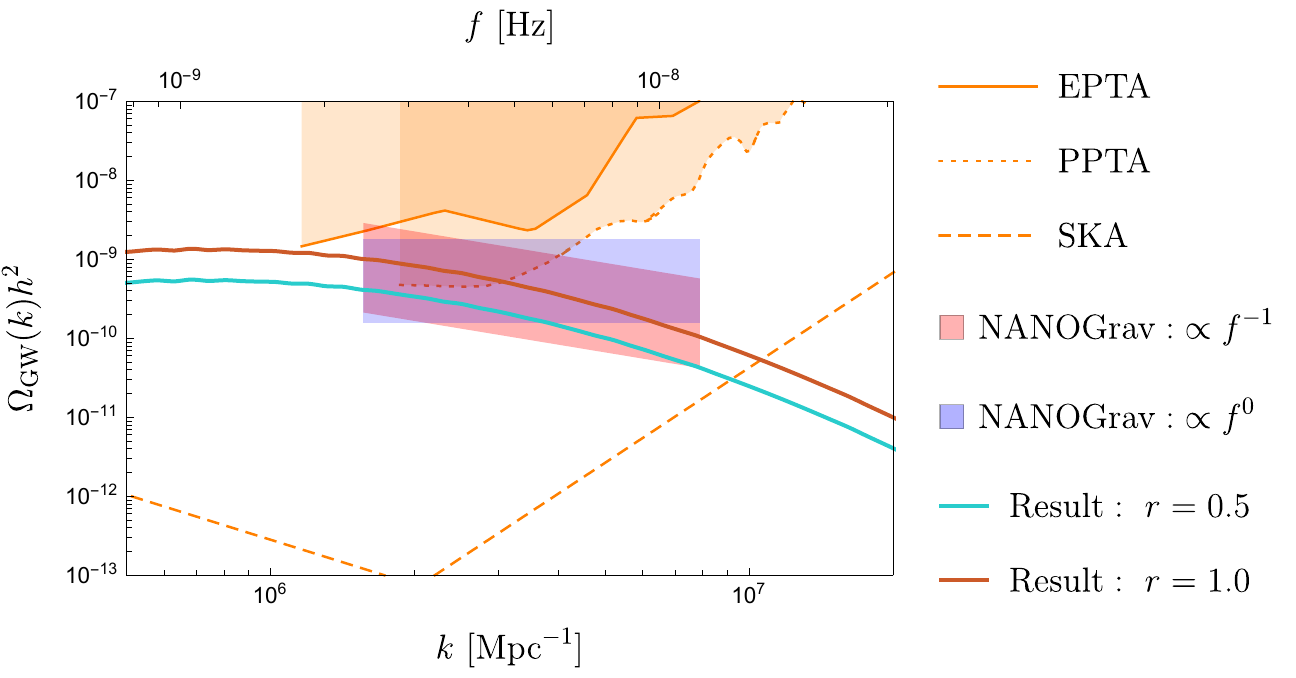}
    \caption{
    The induced GW spectrum and the constraints by PTA experiments.
    We include the contribution of non-Gaussianity by using the factor $Q^{\rm (NL)}_{\mathcal P}$.
    The orange lines are current constraints by EPTA (solid) \cite{Lentati:2015qwp}  and PPTA (Dotted)\cite{Shannon:2015ect}, and future sensitivity by SKA (Dashed)~\cite{Moore:2014lga}.
    We show the reported NANOGrav signal with $\gamma = 5$ (blue region) and $\gamma = 6$ (pink region) with 2-$\sigma$ uncertainty (see Eq.~\eqref{eq_nanograv_GW}).
    }
    \label{fig_gw}
\end{figure}

The estimated GW spectra for $r=0.5$ (blue line) and $r=1.0$ (orange line) are shown in Fig.~\ref{fig_gw}.
The GW spectrum observed by the NANOGrav experiment
is fitted by the power-law spectrum around $f\sim 10^{-8}$, 
\begin{align}
	\Omega_{\rm GW}(f) h^2 = 
		\frac{2\pi^2}{3} \frac{h^2f^2}{H_0^2}  
	h_c^2(f)
	=
	A_{\Omega}  f^{5-\gamma} ,
	\label{eq_nanograv_GW}
 \end{align}
 where $\gamma$ is the tilt of the spectrum.
 In Fig.~\ref{fig_gw}, we show the observed GWs for $\gamma = 5$ and 6 with 2-$\sigma$ uncertainty on $A_{\Omega}$.
 We also plot current constraints by other PTA experiments,  EPTA (solid) \cite{Lentati:2015qwp} and PPTA (dotted)\cite{Shannon:2015ect}, and future sensitivity by SKA (dashed)~\cite{Moore:2014lga}.
It is found that the broad power spectrum of GWs in the present model can explain the reported NANOGrav signal.

\section{Conclusion}
\label{sec_conclusion}

The reported signal by the NANOGrav experiment indicates the various cosmological phenomena like cosmic string, phase transition and PBH formation.
The PBH formation scenario is attractive among them since the reported frequency $f\sim 10^{-8}\Hz$ is close to the scale of the density fluctuations to produce $30\Msol$ PBH, which can explain the binary black holes observed by LIGO-Virgo collaboration.

The typical PBH formation models predict the induced GWs with smaller frequency and larger amplitude compared to the NANOGrav signal.
To avoid this difficulty, one needs to modify the induced GW spectrum by the broad power spectrum and the non-Gaussianity of density fluctuations, which can enhance the PBH formation rate and give a good fit to the NANOGrav signal.
The axion-like curvaton model can achieve the required features.

There are two types of the axion-like curvaton models; in type~I the complex field rolls down toward origin during inflation \cite{Kawasaki:2012wr,Kawasaki:2013xsa,Ando:2017veq,Inomata:2020xad}, and in type~II it rolls down from the origin\cite{Ando:2018nge}.
In this paper, we focused on the type~II model and
chose the appropriate parameters of the potential term, which leads to a broader power spectrum of the density perturbations than that in~\cite{Ando:2018nge}.
As a result it was found that the induced GW spectrum can explain the NANOGrav signal as shown in Fig.~\ref{fig_gw}.
Moreover, the broad power spectrum of the density fluctuation results in the broad mass spectrum of PBHs as shown in Fig.~\ref{fig_pbhdist}, which can be tested by the accumulation of the binary merger observations.
Our model predicts large local-type non-Gauusianity which can be probed through observation of GWs.
The spectrum of induced gravitational waves is a useful tool to distinguish our model from others.
For example, cosmic string and type-I axion-like curvaton models gererally predict much broader GW spectra than our model.

\section*{Acknowledgment}
This work was supported by JSPS KAKENHI Grant Nos. 17H01131 (M.K.), 17K05434 (M.K.), 20H05851 (M.K.), 21K03567(M.K.), JP19J21974 (H.N.), Advanced Leading Graduate Course for Photon Science (H.N.), and 
World Premier International Research Center
Initiative (WPI Initiative), MEXT, Japan (M.K.).


\bibliographystyle{JHEP}
\bibliography{Ref.bib}

\providecommand{\href}[2]{#2}\begingroup\raggedright\begin{thebibliography}{10}

\bibitem{Arzoumanian:2020vkk}
{\scshape NANOGrav} collaboration, Z.~Arzoumanian et~al., \emph{{The NANOGrav
  12.5-year Data Set: Search For An Isotropic Stochastic Gravitational-Wave
  Background}},  \href{https://arxiv.org/abs/2009.04496}{{\ttfamily
  2009.04496}}.

\bibitem{Ellis:2020ena}
J.~Ellis and M.~Lewicki, \emph{{Cosmic String Interpretation of NANOGrav Pulsar
  Timing Data}},  \href{https://arxiv.org/abs/2009.06555}{{\ttfamily
  2009.06555}}.

\bibitem{Blasi:2020mfx}
S.~Blasi, V.~Brdar and K.~Schmitz, \emph{{Has NANOGrav found first evidence for
  cosmic strings?}},  \href{https://arxiv.org/abs/2009.06607}{{\ttfamily
  2009.06607}}.

\bibitem{Buchmuller:2020lbh}
W.~Buchmuller, V.~Domcke and K.~Schmitz, \emph{{From NANOGrav to LIGO with
  metastable cosmic strings}},
  \href{https://doi.org/10.1016/j.physletb.2020.135914}{\emph{Phys. Lett. B}
  {\bfseries 811} (2020) 135914},
  [\href{https://arxiv.org/abs/2009.10649}{{\ttfamily 2009.10649}}].

\bibitem{Samanta:2020cdk}
R.~Samanta and S.~Datta, \emph{{Gravitational wave complementarity and impact
  of NANOGrav data on gravitational leptogenesis: cosmic strings}},
  \href{https://arxiv.org/abs/2009.13452}{{\ttfamily 2009.13452}}.

\bibitem{Datta:2020bht}
S.~Datta, A.~Ghosal, R.~Samanta and R.~Sinha, \emph{{Baryogenesis from
  ultralight primordial black holes and strong gravitational waves}},
  \href{https://arxiv.org/abs/2012.14981}{{\ttfamily 2012.14981}}.

\bibitem{Ramberg:2020oct}
N.~Ramberg and L.~Visinelli, \emph{{The QCD Axion and Gravitational Waves in
  light of NANOGrav results}},  12, 2020,
  \href{https://arxiv.org/abs/2012.06882}{{\ttfamily 2012.06882}}.

\bibitem{Nakai:2020oit}
Y.~Nakai, M.~Suzuki, F.~Takahashi and M.~Yamada, \emph{{Gravitational Waves and
  Dark Radiation from Dark Phase Transition: Connecting NANOGrav Pulsar Timing
  Data and Hubble Tension}},
  \href{https://arxiv.org/abs/2009.09754}{{\ttfamily 2009.09754}}.

\bibitem{Addazi:2020zcj}
A.~Addazi, Y.-F. Cai, Q.~Gan, A.~Marciano and K.~Zeng, \emph{{NANOGrav results
  and Dark First Order Phase Transitions}},
  \href{https://arxiv.org/abs/2009.10327}{{\ttfamily 2009.10327}}.

\bibitem{Neronov:2020qrl}
A.~Neronov, A.~Roper~Pol, C.~Caprini and D.~Semikoz, \emph{{NANOGrav signal
  from MHD turbulence at QCD phase transition in the early universe}},
  \href{https://arxiv.org/abs/2009.14174}{{\ttfamily 2009.14174}}.

\bibitem{Li:2021qer}
S.-L. Li, L.~Shao, P.~Wu and H.~Yu, \emph{{NANOGrav Signal from First-Order
  Confinement/Deconfinement Phase Transition in Different QCD Matters}},
  \href{https://arxiv.org/abs/2101.08012}{{\ttfamily 2101.08012}}.

\bibitem{Barman:2020jrf}
B.~Barman, A.~Dutta~Banik and A.~Paul, \emph{{Implications of NANOGrav results
  and UV freeze-in in a fast-expanding Universe}},
  \href{https://arxiv.org/abs/2012.11969}{{\ttfamily 2012.11969}}.

\bibitem{Vaskonen:2020lbd}
V.~Vaskonen and H.~Veerm\"ae, \emph{{Did NANOGrav see a signal from primordial
  black hole formation?}},  \href{https://arxiv.org/abs/2009.07832}{{\ttfamily
  2009.07832}}.

\bibitem{DeLuca:2020agl}
V.~De~Luca, G.~Franciolini and A.~Riotto, \emph{{NANOGrav Hints to Primordial
  Black Holes as Dark Matter}},
  \href{https://arxiv.org/abs/2009.08268}{{\ttfamily 2009.08268}}.

\bibitem{Inomata:2020xad}
K.~Inomata, M.~Kawasaki, K.~Mukaida and T.~T. Yanagida, \emph{{NANOGrav results
  and LIGO-Virgo primordial black holes in axion-like curvaton model}},
  \href{https://arxiv.org/abs/2011.01270}{{\ttfamily 2011.01270}}.

\bibitem{Kohri:2020qqd}
K.~Kohri and T.~Terada, \emph{{Solar-Mass Primordial Black Holes Explain
  NANOGrav Hint of Gravitational Waves}},
  \href{https://doi.org/10.1016/j.physletb.2020.136040}{\emph{Phys. Lett. B}
  {\bfseries 813} (2021) 136040},
  [\href{https://arxiv.org/abs/2009.11853}{{\ttfamily 2009.11853}}].

\bibitem{Domenech:2020ers}
G.~Dom\`enech and S.~Pi, \emph{{NANOGrav Hints on Planet-Mass Primordial Black
  Holes}},  \href{https://arxiv.org/abs/2010.03976}{{\ttfamily 2010.03976}}.

\bibitem{Sugiyama:2020roc}
S.~Sugiyama, V.~Takhistov, E.~Vitagliano, A.~Kusenko, M.~Sasaki and M.~Takada,
  \emph{{Testing Stochastic Gravitational Wave Signals from Primordial Black
  Holes with Optical Telescopes}},
  \href{https://arxiv.org/abs/2010.02189}{{\ttfamily 2010.02189}}.

\bibitem{Braglia:2020eai}
M.~Braglia, D.~K. Hazra, F.~Finelli, G.~F. Smoot, L.~Sriramkumar and A.~A.
  Starobinsky, \emph{{Generating PBHs and small-scale GWs in two-field models
  of inflation}},
  \href{https://doi.org/10.1088/1475-7516/2020/08/001}{\emph{JCAP} {\bfseries
  08} (2020) 001}, [\href{https://arxiv.org/abs/2005.02895}{{\ttfamily
  2005.02895}}].

\bibitem{Bhaumik:2020dor}
N.~Bhaumik and R.~K. Jain, \emph{{Stochastic induced gravitational waves and
  lowest mass limit of primordial black holes with the effects of reheating}},
  \href{https://arxiv.org/abs/2009.10424}{{\ttfamily 2009.10424}}.

\bibitem{Braglia:2020taf}
M.~Braglia, X.~Chen and D.~Kumar~Hazra, \emph{{Probing Primordial Features with
  the Stochastic Gravitational Wave Background}},
  \href{https://doi.org/10.1088/1475-7516/2021/03/005}{\emph{JCAP} {\bfseries
  03} (2021) 005}, [\href{https://arxiv.org/abs/2012.05821}{{\ttfamily
  2012.05821}}].

\bibitem{Atal:2020yic}
V.~Atal, A.~Sanglas and N.~Triantafyllou, \emph{{NANOGrav signal as mergers of
  Stupendously Large Primordial Black Holes}},
  \href{https://arxiv.org/abs/2012.14721}{{\ttfamily 2012.14721}}.

\bibitem{Vagnozzi:2020gtf}
S.~Vagnozzi, \emph{{Implications of the NANOGrav results for inflation}},
  \href{https://doi.org/10.1093/mnrasl/slaa203}{\emph{Mon. Not. Roy. Astron.
  Soc.} {\bfseries 502} (2021) L11},
  [\href{https://arxiv.org/abs/2009.13432}{{\ttfamily 2009.13432}}].

\bibitem{Bhattacharya:2020lhc}
S.~Bhattacharya, S.~Mohanty and P.~Parashari, \emph{{Implications of the
  NANOGrav result on primordial gravitational waves in nonstandard
  cosmologies}},  \href{https://arxiv.org/abs/2010.05071}{{\ttfamily
  2010.05071}}.

\bibitem{Kuroyanagi:2020sfw}
S.~Kuroyanagi, T.~Takahashi and S.~Yokoyama, \emph{{Blue-tilted inflationary
  tensor spectrum and reheating in the light of NANOGrav results}},
  \href{https://doi.org/10.1088/1475-7516/2021/01/071}{\emph{JCAP} {\bfseries
  01} (2021) 071}, [\href{https://arxiv.org/abs/2011.03323}{{\ttfamily
  2011.03323}}].

\bibitem{LIGOScientific:2018jsj}
{\scshape LIGO Scientific, Virgo} collaboration, B.~P. Abbott et~al.,
  \emph{{Binary Black Hole Population Properties Inferred from the First and
  Second Observing Runs of Advanced LIGO and Advanced Virgo}},
  \href{https://arxiv.org/abs/1811.12940}{{\ttfamily 1811.12940}}.

\bibitem{Bird:2016dcv}
S.~Bird, I.~Cholis, J.~B. Muñoz, Y.~Ali-Haïmoud, M.~Kamionkowski, E.~D.
  Kovetz et~al., \emph{{Did LIGO detect dark matter?}},
  \href{https://doi.org/10.1103/PhysRevLett.116.201301}{\emph{Phys. Rev. Lett.}
  {\bfseries 116} (2016) 201301},
  [\href{https://arxiv.org/abs/1603.00464}{{\ttfamily 1603.00464}}].

\bibitem{Clesse:2016vqa}
S.~Clesse and J.~García-Bellido, \emph{{The clustering of massive Primordial
  Black Holes as Dark Matter: measuring their mass distribution with Advanced
  LIGO}}, \href{https://doi.org/10.1016/j.dark.2016.10.002}{\emph{Phys. Dark
  Univ.} {\bfseries 15} (2017) 142--147},
  [\href{https://arxiv.org/abs/1603.05234}{{\ttfamily 1603.05234}}].

\bibitem{Sasaki:2016jop}
M.~Sasaki, T.~Suyama, T.~Tanaka and S.~Yokoyama, \emph{{Primordial Black Hole
  Scenario for the Gravitational-Wave Event GW150914}},
  \href{https://doi.org/10.1103/PhysRevLett.117.061101}{\emph{Phys. Rev. Lett.}
  {\bfseries 117} (2016) 061101},
  [\href{https://arxiv.org/abs/1603.08338}{{\ttfamily 1603.08338}}].

\bibitem{Inomata:2017vxo}
K.~Inomata, M.~Kawasaki, K.~Mukaida and T.~T. Yanagida, \emph{{Double inflation
  as a single origin of primordial black holes for all dark matter and LIGO
  observations}}, \href{https://doi.org/10.1103/PhysRevD.97.043514}{\emph{Phys.
  Rev.} {\bfseries D97} (2018) 043514},
  [\href{https://arxiv.org/abs/1711.06129}{{\ttfamily 1711.06129}}].

\bibitem{1975ApJ...201....1C}
B.~J. Carr, \emph{{The Primordial black hole mass spectrum}},
  \href{https://doi.org/10.1086/153853}{\emph{Astrophys. J.} {\bfseries 201}
  (1975) 1--19}.

\bibitem{Saito:2008jc}
R.~Saito and J.~Yokoyama, \emph{{Gravitational wave background as a probe of
  the primordial black hole abundance}},
  \href{https://doi.org/10.1103/PhysRevLett.102.161101}{\emph{Phys. Rev. Lett.}
  {\bfseries 102} (2009) 161101},
  [\href{https://arxiv.org/abs/0812.4339}{{\ttfamily 0812.4339}}].

\bibitem{saito2010gravitational}
R.~Saito and J.~Yokoyama, \emph{Gravitational-wave constraints on the abundance
  of primordial black holes}, {\emph{Progress of theoretical physics}
  {\bfseries 123} (2010) 867--886}.

\bibitem{Unal:2020mts}
C.~Unal, E.~D. Kovetz and S.~P. Patil, \emph{{Multi-messenger Probes of
  Inflationary Fluctuations and Primordial Black Holes}},
  \href{https://arxiv.org/abs/2008.11184}{{\ttfamily 2008.11184}}.

\bibitem{Kawasaki:2012wr}
M.~Kawasaki, N.~Kitajima and T.~T. Yanagida, \emph{{Primordial black hole
  formation from an axionlike curvaton model}},
  \href{https://doi.org/10.1103/PhysRevD.87.063519}{\emph{Phys. Rev.}
  {\bfseries D87} (2013) 063519},
  [\href{https://arxiv.org/abs/1207.2550}{{\ttfamily 1207.2550}}].

\bibitem{Kawasaki:2013xsa}
M.~Kawasaki, N.~Kitajima and S.~Yokoyama, \emph{{Gravitational waves from a
  curvaton model with blue spectrum}},
  \href{https://doi.org/10.1088/1475-7516/2013/08/042}{\emph{JCAP} {\bfseries
  1308} (2013) 042}, [\href{https://arxiv.org/abs/1305.4464}{{\ttfamily
  1305.4464}}].

\bibitem{Ando:2017veq}
K.~Ando, K.~Inomata, M.~Kawasaki, K.~Mukaida and T.~T. Yanagida,
  \emph{{Primordial Black Holes for the LIGO Events in the Axion-like Curvaton
  Model}},  \href{https://arxiv.org/abs/1711.08956}{{\ttfamily 1711.08956}}.

\bibitem{Ando:2018nge}
K.~Ando, M.~Kawasaki and H.~Nakatsuka, \emph{{Formation of primordial black
  holes in an axionlike curvaton model}},
  \href{https://doi.org/10.1103/PhysRevD.98.083508}{\emph{Phys. Rev. D}
  {\bfseries 98} (2018) 083508},
  [\href{https://arxiv.org/abs/1805.07757}{{\ttfamily 1805.07757}}].

\bibitem{tHooft:1979rat}
G.~'t~Hooft, \emph{{Naturalness, chiral symmetry, and spontaneous chiral
  symmetry breaking}},
  \href{https://doi.org/10.1007/978-1-4684-7571-5_9}{\emph{NATO Sci. Ser. B}
  {\bfseries 59} (1980) 135--157}.

\bibitem{Unal:2018yaa}
C.~Unal, \emph{{Imprints of Primordial Non-Gaussianity on Gravitational Wave
  Spectrum}}, \href{https://doi.org/10.1103/PhysRevD.99.041301}{\emph{Phys.
  Rev. D} {\bfseries 99} (2019) 041301},
  [\href{https://arxiv.org/abs/1811.09151}{{\ttfamily 1811.09151}}].

\bibitem{Cai:2018dig}
R.-g. Cai, S.~Pi and M.~Sasaki, \emph{{Gravitational Waves Induced by
  non-Gaussian Scalar Perturbations}},
  \href{https://doi.org/10.1103/PhysRevLett.122.201101}{\emph{Phys. Rev. Lett.}
  {\bfseries 122} (2019) 201101},
  [\href{https://arxiv.org/abs/1810.11000}{{\ttfamily 1810.11000}}].

\bibitem{0004-637X-758-2-76}
J.~Chluba, A.~L. Erickcek and I.~Ben-Dayan, \emph{Probing the inflaton:
  Small-scale power spectrum constraints from measurements of the cosmic
  microwave background energy spectrum}, {\emph{The Astrophysical Journal}
  {\bfseries 758} (2012) 76}.

\bibitem{Fixsen:1996nj}
D.~J. Fixsen, E.~S. Cheng, J.~M. Gales, J.~C. Mather, R.~A. Shafer and E.~L.
  Wright, \emph{{The Cosmic Microwave Background spectrum from the full COBE
  FIRAS data set}}, \href{https://doi.org/10.1086/178173}{\emph{Astrophys. J.}
  {\bfseries 473} (1996) 576},
  [\href{https://arxiv.org/abs/astro-ph/9605054}{{\ttfamily
  astro-ph/9605054}}].

\bibitem{Inomata:2016uip}
K.~Inomata, M.~Kawasaki and Y.~Tada, \emph{{Revisiting constraints on small
  scale perturbations from big-bang nucleosynthesis}},
  \href{https://doi.org/10.1103/PhysRevD.94.043527}{\emph{Phys. Rev.}
  {\bfseries D94} (2016) 043527},
  [\href{https://arxiv.org/abs/1605.04646}{{\ttfamily 1605.04646}}].

\bibitem{Shibata:1999zs}
M.~Shibata and M.~Sasaki, \emph{{Black hole formation in the Friedmann
  universe: Formulation and computation in numerical relativity}},
  \href{https://doi.org/10.1103/PhysRevD.60.084002}{\emph{Phys. Rev. D}
  {\bfseries 60} (1999) 084002},
  [\href{https://arxiv.org/abs/gr-qc/9905064}{{\ttfamily gr-qc/9905064}}].

\bibitem{Kawasaki:2019mbl}
M.~Kawasaki and H.~Nakatsuka, \emph{{Effect of nonlinearity between density and
  curvature perturbations on the primordial black hole formation}},
  \href{https://doi.org/10.1103/PhysRevD.99.123501}{\emph{Phys. Rev. D}
  {\bfseries 99} (2019) 123501},
  [\href{https://arxiv.org/abs/1903.02994}{{\ttfamily 1903.02994}}].

\bibitem{Ando:2018qdb}
K.~Ando, K.~Inomata and M.~Kawasaki, \emph{{Primordial black holes and
  uncertainties on choice of window function}},
  \href{https://arxiv.org/abs/1802.06393}{{\ttfamily 1802.06393}}.

\bibitem{Young:2013oia}
S.~Young and C.~T. Byrnes, \emph{{Primordial black holes in non-Gaussian
  regimes}}, \href{https://doi.org/10.1088/1475-7516/2013/08/052}{\emph{JCAP}
  {\bfseries 1308} (2013) 052},
  [\href{https://arxiv.org/abs/1307.4995}{{\ttfamily 1307.4995}}].

\bibitem{Byrnes:2012yx}
C.~T. Byrnes, E.~J. Copeland and A.~M. Green, \emph{{Primordial black holes as
  a tool for constraining non-Gaussianity}},
  \href{https://doi.org/10.1103/PhysRevD.86.043512}{\emph{Phys. Rev.}
  {\bfseries D86} (2012) 043512},
  [\href{https://arxiv.org/abs/1206.4188}{{\ttfamily 1206.4188}}].

\bibitem{Young:2015cyn}
S.~Young, D.~Regan and C.~T. Byrnes, \emph{{Influence of large local and
  non-local bispectra on primordial black hole abundance}},
  \href{https://doi.org/10.1088/1475-7516/2016/02/029}{\emph{JCAP} {\bfseries
  1602} (2016) 029}, [\href{https://arxiv.org/abs/1512.07224}{{\ttfamily
  1512.07224}}].

\bibitem{Allsman:2000kg}
{\scshape Macho} collaboration, R.~A. Allsman et~al., \emph{{MACHO project
  limits on black hole dark matter in the 1-30 solar mass range}},
  \href{https://doi.org/10.1086/319636}{\emph{Astrophys. J.} {\bfseries 550}
  (2001) L169}, [\href{https://arxiv.org/abs/astro-ph/0011506}{{\ttfamily
  astro-ph/0011506}}].

\bibitem{Tisserand:2006zx}
{\scshape EROS-2} collaboration, P.~Tisserand et~al., \emph{{Limits on the
  Macho Content of the Galactic Halo from the EROS-2 Survey of the Magellanic
  Clouds}}, \href{https://doi.org/10.1051/0004-6361:20066017}{\emph{Astron.
  Astrophys.} {\bfseries 469} (2007) 387--404},
  [\href{https://arxiv.org/abs/astro-ph/0607207}{{\ttfamily
  astro-ph/0607207}}].

\bibitem{Wyrzykowski:2011tr}
L.~Wyrzykowski et~al., \emph{{The OGLE View of Microlensing towards the
  Magellanic Clouds. IV. OGLE-III SMC Data and Final Conclusions on MACHOs}},
  \href{https://doi.org/10.1111/j.1365-2966.2011.19243.x}{\emph{Mon. Not. Roy.
  Astron. Soc.} {\bfseries 416} (2011) 2949},
  [\href{https://arxiv.org/abs/1106.2925}{{\ttfamily 1106.2925}}].

\bibitem{Serpico:2020ehh}
P.~D. Serpico, V.~Poulin, D.~Inman and K.~Kohri, \emph{{Cosmic microwave
  background bounds on primordial black holes including dark matter halo
  accretion}},
  \href{https://doi.org/10.1103/PhysRevResearch.2.023204}{\emph{Phys. Rev.
  Res.} {\bfseries 2} (2020) 023204},
  [\href{https://arxiv.org/abs/2002.10771}{{\ttfamily 2002.10771}}].

\bibitem{Garcia-Bellido:2017aan}
J.~Garcia-Bellido, M.~Peloso and C.~Unal, \emph{{Gravitational Wave signatures
  of inflationary models from Primordial Black Hole Dark Matter}},
  \href{https://doi.org/10.1088/1475-7516/2017/09/013}{\emph{JCAP} {\bfseries
  09} (2017) 013}, [\href{https://arxiv.org/abs/1707.02441}{{\ttfamily
  1707.02441}}].

\bibitem{Lentati:2015qwp}
L.~Lentati et~al., \emph{{European Pulsar Timing Array Limits On An Isotropic
  Stochastic Gravitational-Wave Background}},
  \href{https://doi.org/10.1093/mnras/stv1538}{\emph{Mon. Not. Roy. Astron.
  Soc.} {\bfseries 453} (2015) 2576--2598},
  [\href{https://arxiv.org/abs/1504.03692}{{\ttfamily 1504.03692}}].

\bibitem{Shannon:2015ect}
R.~M. Shannon et~al., \emph{{Gravitational waves from binary supermassive black
  holes missing in pulsar observations}},
  \href{https://doi.org/10.1126/science.aab1910}{\emph{Science} {\bfseries 349}
  (2015) 1522--1525}, [\href{https://arxiv.org/abs/1509.07320}{{\ttfamily
  1509.07320}}].

\bibitem{Moore:2014lga}
C.~J. Moore, R.~H. Cole and C.~P.~L. Berry, \emph{{Gravitational-wave
  sensitivity curves}},
  \href{https://doi.org/10.1088/0264-9381/32/1/015014}{\emph{Class. Quant.
  Grav.} {\bfseries 32} (2015) 015014},
  [\href{https://arxiv.org/abs/1408.0740}{{\ttfamily 1408.0740}}].

\end{thebibliography}\endgroup
\end{document}